\documentclass[conference]{IEEEtran}
\IEEEoverridecommandlockouts
\usepackage[table]{xcolor}
\usepackage{collcell}
\usepackage{pgfmath}
\usepackage{subcaption}
\usepackage{xstring}    
\usepackage{balance}  
\usepackage{tikz}      

\definecolor{low}{RGB}{255,230,128}    
\definecolor{high}{RGB}{ 30, 74,155}   

\def\CORRmin{-0.02}
\def\CORRmax{1.00}

\newcommand\CorrHeat[1]{%
  \IfDecimal{#1}{%
    \pgfmathparse{int(round(100*(#1-\CORRmin)/(\CORRmax-\CORRmin)))}%
    \xdef\Percent{\pgfmathresult}
    \cellcolor{low!\Percent!high}{#1}%
  }{#1}%
}

\newcolumntype{H}{>{\collectcell\CorrHeat}r<{\endcollectcell}}

\usepackage{cite}
\usepackage{amsmath,amssymb,amsfonts}
\usepackage{algorithmic}
\usepackage{graphicx}
\usepackage{textcomp}
\usepackage{xcolor}
\usepackage{longtable}
\usepackage{booktabs}
\def\BibTeX{{\rm B\kern-.05em{\sc i\kern-.025em b}\kern-.08em
    T\kern-.1667em\lower.7ex\hbox{E}\kern-.125emX}}
\begin{document}


\title{On the Limitations of Ray-Tracing for Learning-Based RF Tasks in Urban Environments}

\author{
\IEEEauthorblockN{ 
Armen Manukyan\IEEEauthorrefmark{1}\IEEEauthorrefmark{2}, 
Hrant Khachatrian\IEEEauthorrefmark{1}\IEEEauthorrefmark{2}, 
Edvard Ghukasyan\IEEEauthorrefmark{1}\IEEEauthorrefmark{2},
Theofanis P. Raptis\IEEEauthorrefmark{3}
}
\IEEEauthorblockA{
\IEEEauthorblockA{\IEEEauthorrefmark{1}Yerevan State University, Yerevan, Armenia. Email: \{hrant.khachatrian, edvard.ghukasyan\}@ysu.am}
\IEEEauthorblockA{\IEEEauthorrefmark{2}YerevaNN, Yerevan, Armenia. Email: armen@yerevann.com}
\IEEEauthorrefmark{3}Institute of Informatics and Telematics, National Research Council, Pisa, Italy. Email: theofanis.raptis@iit.cnr.it
}
}

\maketitle
\begin{tikzpicture}[remember picture,overlay]
\node[anchor=south,yshift=10pt] at (current page.south) {\fbox{\parbox{\dimexpr\textwidth-\fboxsep-\fboxrule\relax}{
  \footnotesize{
    \copyright 2026 IEEE. Personal use of this material is permitted.  Permission from IEEE must be obtained for all other uses, in any current or future media, including reprinting/republishing this material for advertising or promotional purposes, creating new collective works, for resale or redistribution to servers or lists, or reuse of any copyrighted component of this work in other works.
  }
}}};
\end{tikzpicture}

\begin{abstract}
We study the realism of \textsc{Sionna}~v1.0.2 ray-tracing for outdoor cellular links in central Rome. 
We use a real measurement set of 1\,664 user-equipments (UEs) and six nominal base-station (BS) sites.
Using these fixed positions we systematically vary the main simulation parameters, including path depth, diffuse/specular/refraction flags, carrier frequency, as well as antenna's properties like its altitude, radiation pattern, and orientation. Simulator fidelity is scored for each base station via Spearman correlation between measured and simulated powers, and by a fingerprint-based k-nearest-neighbor localization algorithm using RSSI-based fingerprints.  
Across all experiments, solver hyper-parameters are having immaterial effect on the chosen metrics. On the contrary, antenna locations and orientations prove decisive. By simple greedy optimization we improve the Spearman correlation by 5\% to 130\% for various base stations, while kNN-based localization error using only simulated data as reference points is decreased by one-third on real-world samples, while staying twice higher than the error with purely real data. Precise geometry and credible antenna models are therefore necessary but not sufficient; faithfully capturing the residual urban noise remains an open challenge for transferable, high-fidelity outdoor RF simulation.

\end{abstract}


\section{Introduction}

The increasing adoption of wireless technologies in dense urban environments has intensified the need for accurate radio propagation modeling \cite{9714331}. While empirical datasets remain the gold standard for evaluating wireless systems, their collection is often expensive, time-consuming, and limited by physical burdens \cite{9682024}. In this context, ray-tracing simulators have emerged as a powerful alternative, enabling the generation of synthetic radio frequency (RF) data under controlled yet realistic conditions \cite{Yun2024}. These simulators promise not only repeatability and scalability but also the potential to overcome the scarcity of large-scale, labeled wireless datasets \cite{KHACHATRIAN2025103696}.

Despite recent advances in simulator realism, a significant gap persists between synthetic and real-world radio data. This simulation-to-reality gap undermines the effectiveness of simulators in critical downstream tasks, such as signal strength prediction, wireless localization or outdoor mapping \cite{10592367}. The fidelity of synthetic radio data hinges on complex simulation parameters—such as the number of reflections or the transmitter antenna model—whose interplay is poorly understood. Although recent frameworks like \textsc{Sionna} ray-tracing \cite{sionna-rt} integrate differentiable ray tracing with state-of-the-art 3D modeling capabilities, little is known about how accurately they replicate the behavior of real urban wireless deployments.

In this paper, we investigate the extent to which synthetic RF data generated by \textsc{Sionna RT} can match real-world measurements in a dense urban setting. We focus on a representative subregion of the publicly available Rome wireless dataset \cite{RomeData}, which features 1664 user equipment (UE) locations and 10 base stations (BS) in a realistic city layout. Using a custom pipeline, we reproduce the 3D urban environment, place the transmitters and receivers accordingly, and generate simulated RF traces under varying simulation configurations. Our contributions are as follows:

\begin{itemize}
\item We present an end-to-end framework to convert real-world UE/BS coordinates into a 3D simulation-ready scene for Sionna RT, integrating publicly available building data.
\item We conduct a detailed parameter sweep on core simulation settings, including number of reflections, scene resolution, and transmitter antenna model.
\item We propose a two-fold evaluation strategy: (i) computing the Spearman correlation between real and synthetic RF values, and (ii) assessing k-nearest neighbor (kNN) localization accuracy using real, synthetic, and hybrid RF fingerprints.
\item We analyze the impact of simulation choices on both signal fidelity and downstream localization, providing practical insights for wireless researchers leveraging ray-tracing tools.
\end{itemize}
The remainder of the paper is organized as follows. Section \ref{sec:rel} reviews some recent related works on radio simulation and synthetic dataset evaluation. Section \ref{sec:met} describes the dataset, simulation framework, and experimental setup. Section \ref{sec:eva} presents our quantitative results and analysis and discusses the implications and limitations of our findings. Finally, Section \ref{sec:con} concludes the paper and outlines future research directions.



\section{Related Work} \label{sec:rel}
Ray-tracing engines, such as \textsc{Sionna RT} \cite{sionna-rt}, have been increasingly used to create high-fidelity digital counterparts of wireless channels. For outdoor urban scenarios, recent works have demonstrated that a fairly coarse 3D city model can reproduce large-scale signal trends with reasonable accuracy, so long as critical details (like the actual antenna radiation patterns) are incorporated \cite{10827549}. In fact, using measured antenna patterns was shown to cut typical prediction errors down to about 4 dB, highlighting that an accurate hardware model is as important as an accurate environment model \cite{Schott2023EUCAP}.
For indoor settings as well, achieving simulation fidelity is even more demanding, especially as carrier frequencies climb into the sub‑THz and THz ranges.  Early mmWave validation campaigns showed that ray tracers capture the strongest line‑of‑sight and first‑order reflections but still miss a tail of weak, high‑order multipath, leading to residual errors of a few dB or nanoseconds in power‐delay spectra~\cite{Di2024EUCAP}. 

\begin{figure*}
  \centering
  \begin{subfigure}[b]{0.30\textwidth}
    \centering
    \includegraphics[width=\linewidth]{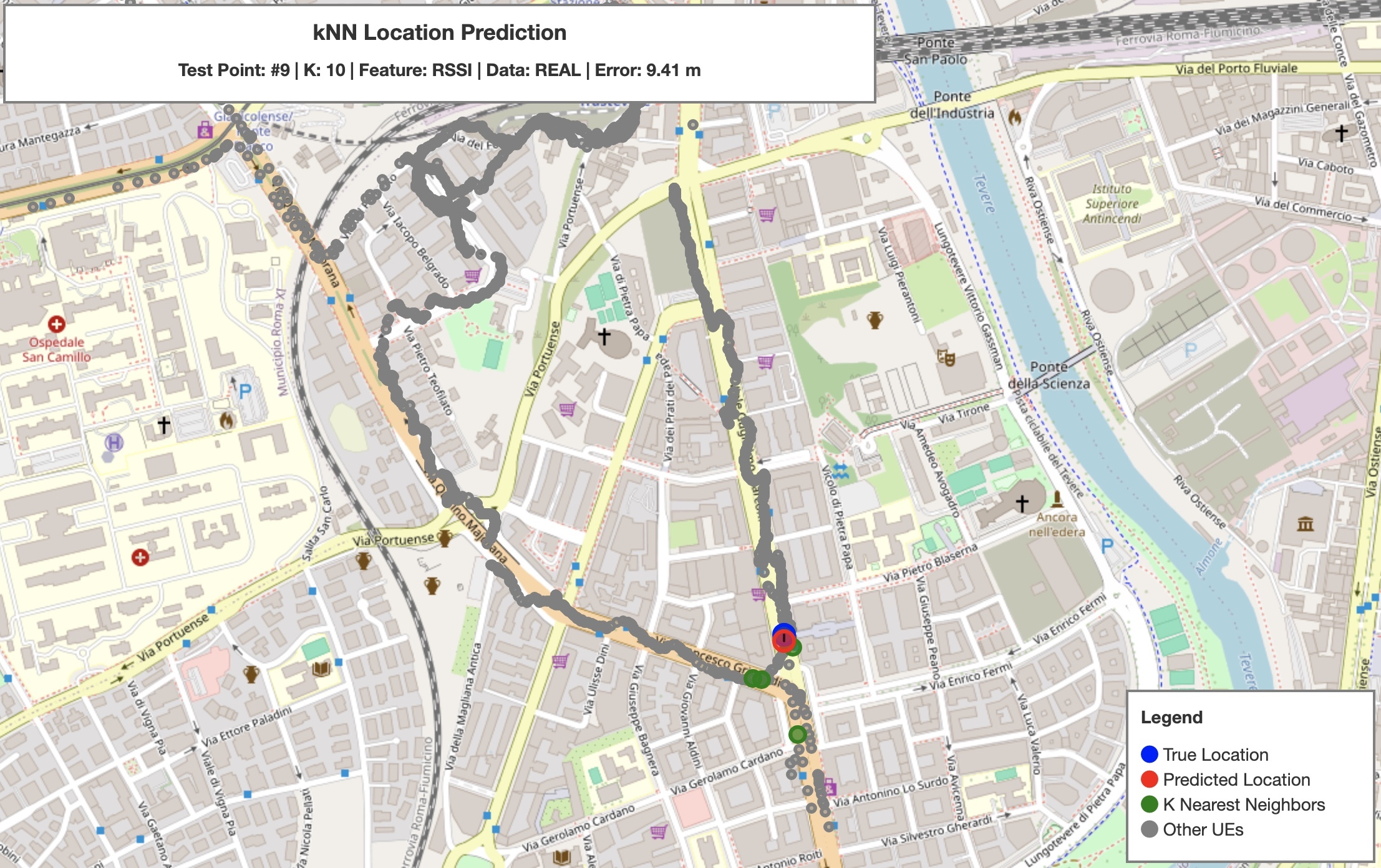}
    \caption{An example with a very low error}
    \label{fig:knn-9}
  \end{subfigure}\hfill
  \begin{subfigure}[b]{0.30\textwidth}
    \centering
    \includegraphics[width=\linewidth]{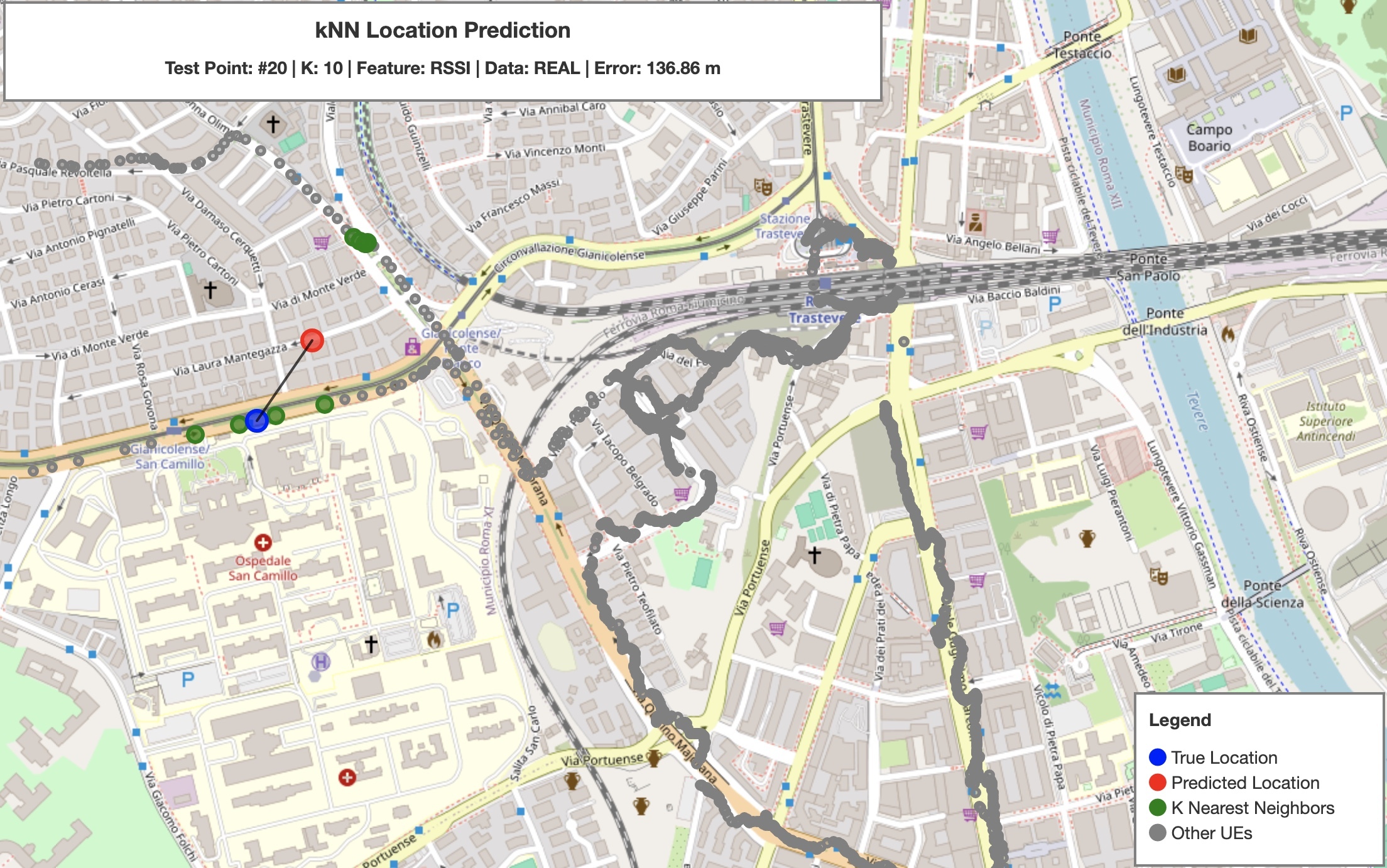}
    \caption{A typical example}
    \label{fig:knn-138}
  \end{subfigure}\hfill
  \begin{subfigure}[b]{0.30\textwidth}
    \centering
    \includegraphics[width=\linewidth]{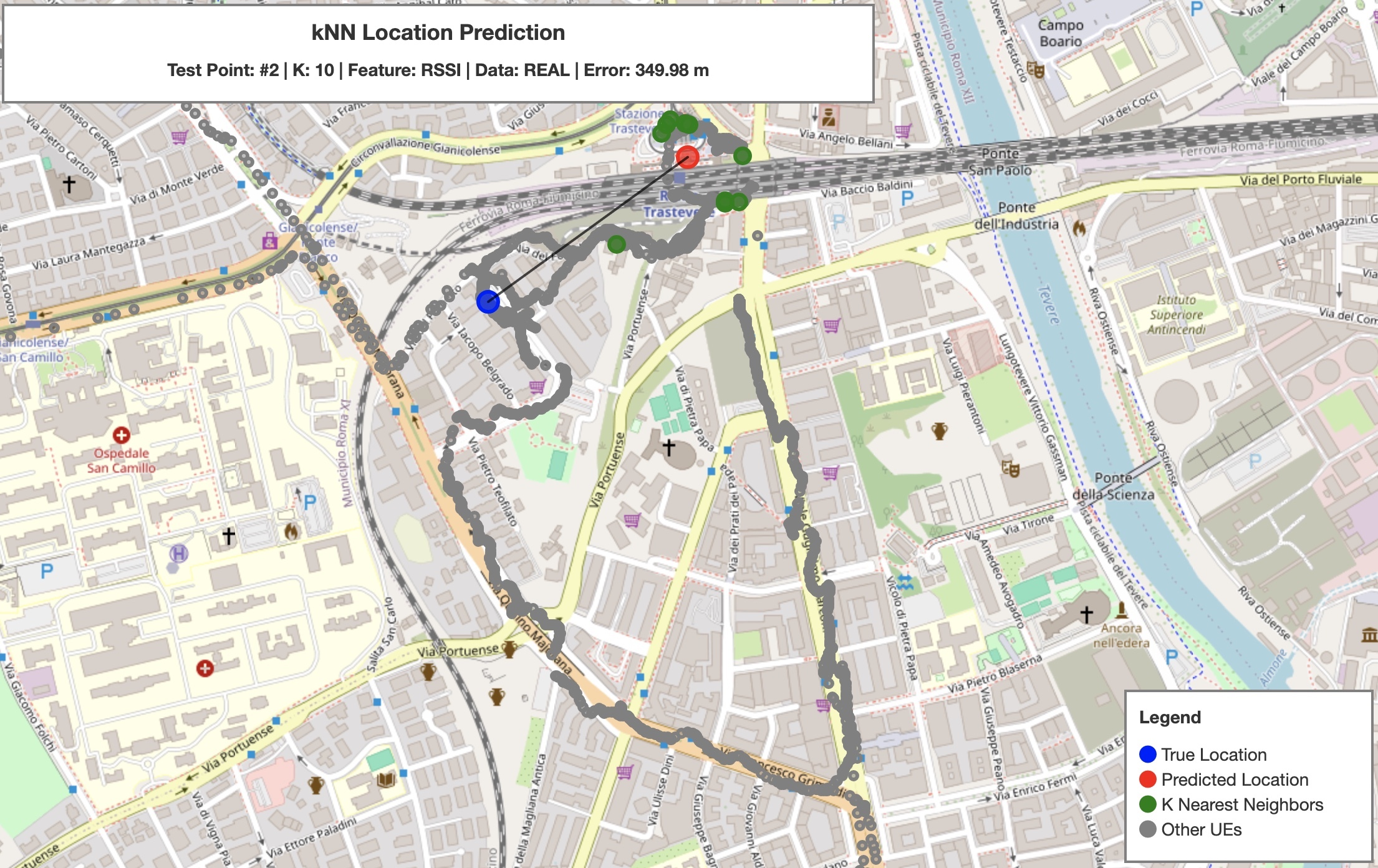}
    \caption{An example with a huge error}
    \label{fig:knn-350}
  \end{subfigure}

  \caption{Three examples of kNN predictions in R $\rightarrow$ R setting. All fingerprints are calculated using real RSSI values. From left to right the images show the easiest, typical, and the hardest cases of the kNN algorithm.}
  \label{fig:knn}
\end{figure*}

Such works however, do not typically take into account the realistic option of invloving both a ray tracer and the real-world counterpart for the validation of the fidelity. The most relevant work to this paper in the state of the art was presented in last year's IEEE GlobeCom by Aksoy \emph{et al.} In that paper, the authors extend that observation to 300GHz by benchmarking \textsc{Sionna}, \emph{WirelessInSite}, and an smale-scale in‑house tracer in a furnished conference room with angle‑resolved measurements~\cite{Aksoy2024THzRT}. They show that default scattering and material models underestimate diffuse energy and angular spread, whereas incorporating Rayleigh–Rice scattering, accurate dielectric parameters, and atmospheric attenuation can cut path‑loss error from \(\sim\!10\)dB to \(\lesssim\!2\)dB. Their findings already hint that high‑frequency realism hinges on refined interaction mechanisms and scene‑specific calibration—constraints. Another related paper is \cite{10666097}, the authors of which use both real and simulated data for a small-scale case, On the contrary to those papers, in our paper, we not only take into account a large-scale (compared to both\cite{Aksoy2024THzRT} and \cite{10666097}) city-wide setting, but also the fundamentally different case (compared to \cite{Aksoy2024THzRT}) of outdoor environment.

\section{Methodology} \label{sec:met}




\subsection{Real-world data}

The real-world measurements used in our experiments are derived from a publicly available dataset~\cite{RomeData} collected in Rome, Italy, covering extensive 4G and 5G network measurements across low-to-mid frequency bands (0.8–4 GHz). From this dataset, we identified a spatial subset containing a large concentration of transmitters and receivers. The plan was to perform hundreds of simulations, so we had to choose a relatively small area. 

The selected region was defined by the minimal bounding rectangle encompassing a dense collection of 1664 UEs associated initially with 10 BS sites. 

\subsection{Simulation environment}

The BS positions and UE coordinates were imported directly into the \textsc{Sionna} ray-tracing simulator~\cite{sionna-rt}. We used version 1.0.2~\cite{sionna102} which brings improvements in speed and memory efficiency. 

We used OpenStreetMap API to retrieve buildings in the selected area, and created a 3D scene in a Mitsuba XML format. Two‑dimensional building footprints were converted into a lightweight 3‑D city model by extruding each footprint vertically by 10m, thereby representing every building as a right prism of constant height. OSM API reports number of floors for a small number of buildings, in which case we set the height of the building equal to 3 times the number of floors. There is also a large rectangular surface below the buildings that acts as a ground. All 3D objects are built from the same concrete material.

We performed a large number of ray-tracing simulations using this scene by varying key simulation parameters and by varying BS and UE configurations, to identify which parameters are the simulations most sensitive to.

\subsection{Evaluation}\label{sec:eval}
For every simulation we compute RSSI values for each BS-UE pair, and perform two kinds of evaluation. First, we compute Spearman correlation between the real and simulated RSSI values for each base station. We chose Spearman over Pearson because there could be shifts between real world and simulated environments by a constant term due to the differences in transmitter and receiver gains and other factors that are hard to control. 

The first evaluation, while important for estimating the simulation-to-real gap, does not tell the whole story. In practice we need simulations for generating synthetic data for downstream applications. One of those applications is the localization of the UE. The second kind of evaluation we designed measures the utility of the simulated data for this particular application. 

We perform a device localization experiment by utlizing RSSI-based fingerprints using k-nearest-neighbor (kNN) algorithm. This closely follows the localization experiment in \cite{RomeData}. For each UE location we compute a fingerprint of that location by taking a vector of $M$ RSSI values, where $M$ is the number of base stations. We split the set of UEs into training and test sets. For each of the UEs from the test set we select its $k=10$ nearest fingerprints from the training set, and take the mean geographic location of those $k$ points as the predicted location of the given UE. The error of the prediction for that UE is defined as the Euclidean distance between the ground-truth and predicted locations, measured in meters.

For each UE location we have two fingerprints: RSSI values taken from the ground truth dataset, and from the simulation. Both training and test set fingerprints can be computed using either sources of RSSI values, which gives us four combinations and, hence, four metrics. 

When both training and test fingerprints are constructed using real-world data, the resulting error metric is the baseline localization performance one would get without simulations  
[\textbf{R$\rightarrow$R}]. This is the closest metric to the ones reported in \cite{RomeData}. The average prediction error here is 118m. Fig. \ref{fig:knn} shows three indicative examples on how the kNN works and how it fails in our subset.  

When both training and test fingerprints come from simulations, the resulting metric shows the localization error inside the simulator. We expect this error to be lower than the previous one, as the simulation is usually less noisy and less dependent on environmental factors that are hard to capture. [\textbf{S$\rightarrow$S}]

For measuring the usefulness of the simulated data in downstream applications we are more interested in the \textit{mixed} setting, when the test set fingerprints are based on real-world RSSI values, while the training set fingerprints come from simulations. This is representing a scenario when the device is in the part of the city which was not covered by earlier data collection activities, so there are no nearby reference points. The error metric for this setting shows how reliably one can use simulated fingerprints for localizing a device in the real world [\textbf{S$\rightarrow$R}].

\begin{figure}
    \centering
    \includegraphics[width=0.8\linewidth]{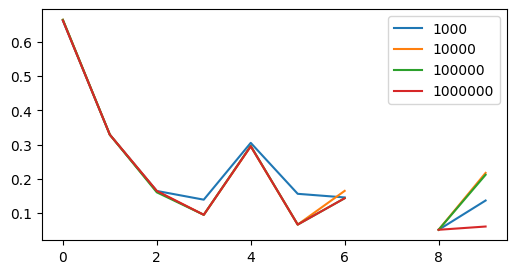}
    \caption{Sensitivity to the \texttt{samples\_per\_src} parameter of the \texttt{PathSolver} class of Sionna simulator. x-axis corresponds to base stations. y-axis is the Spearman correlation between simulated and real RSSI values. While $10^3$ (blue) gives higher correlations, larger values correspond to more accurate simulations, so we choose $10^6$ as the default value for our further experiments.}
    \label{fig:samples-per-src}
\end{figure}

For the sake of completeness, we also report the error when the training set fingerprints use real data while the test set fingerprints use simulated data [\textbf{R$\rightarrow$S}].



\section{Results and Insights} \label{sec:eva}

\begin{figure*}[t!]
  \centering
  \begin{subfigure}[b]{0.32\textwidth}
    \centering
    \includegraphics[width=\linewidth]{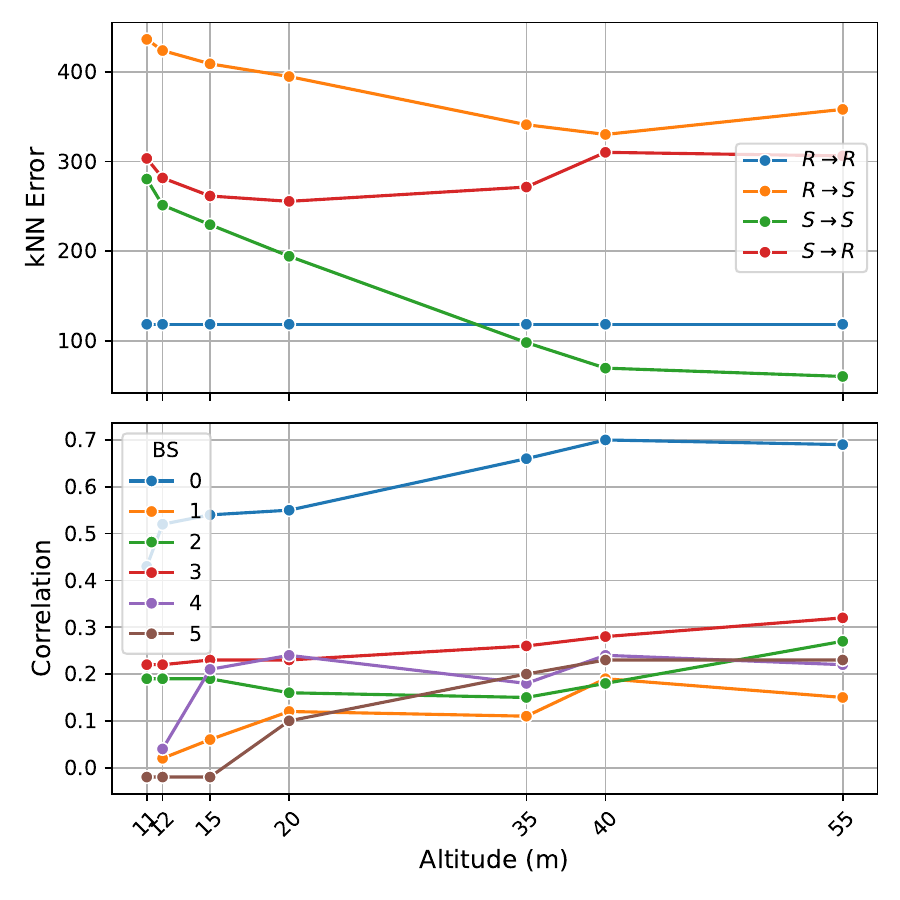}
    \caption{BS altitude}
    \label{fig:altitude}
  \end{subfigure}\hfill
  \begin{subfigure}[b]{0.32\textwidth}
    \centering
    \includegraphics[width=\linewidth]{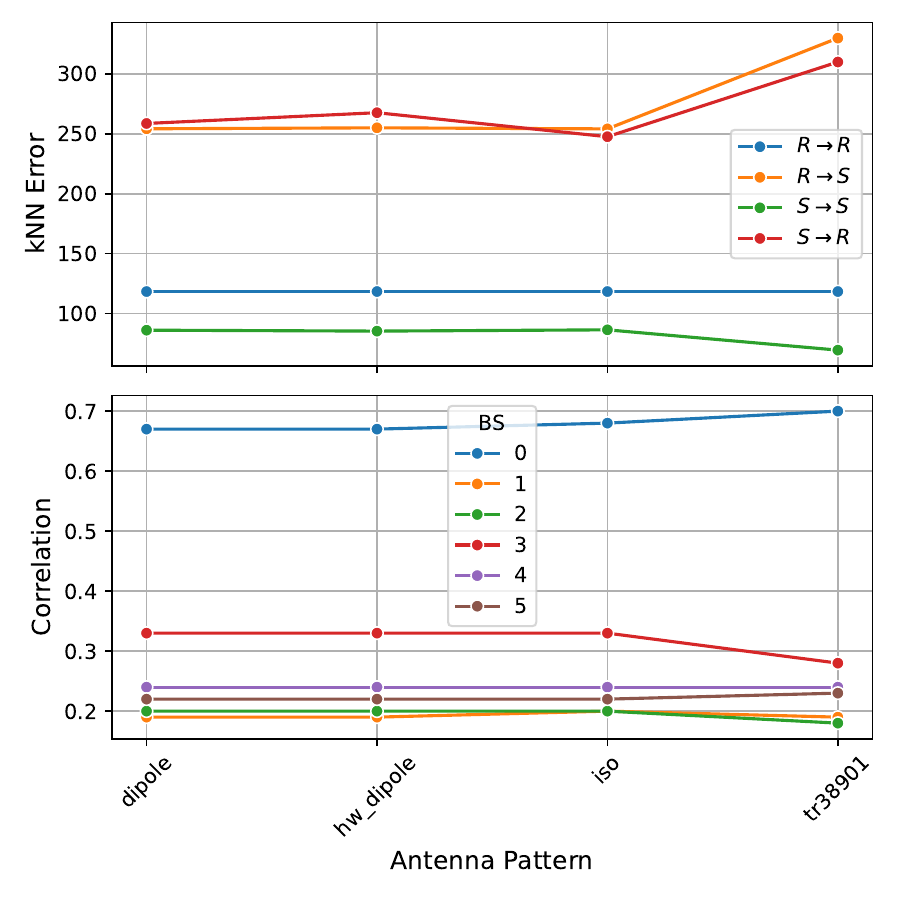}
    \caption{BS radiation pattern}
    \label{fig:pattern}
  \end{subfigure}
  \begin{subfigure}[b]{0.32\textwidth}
    \centering
    \includegraphics[width=\linewidth]{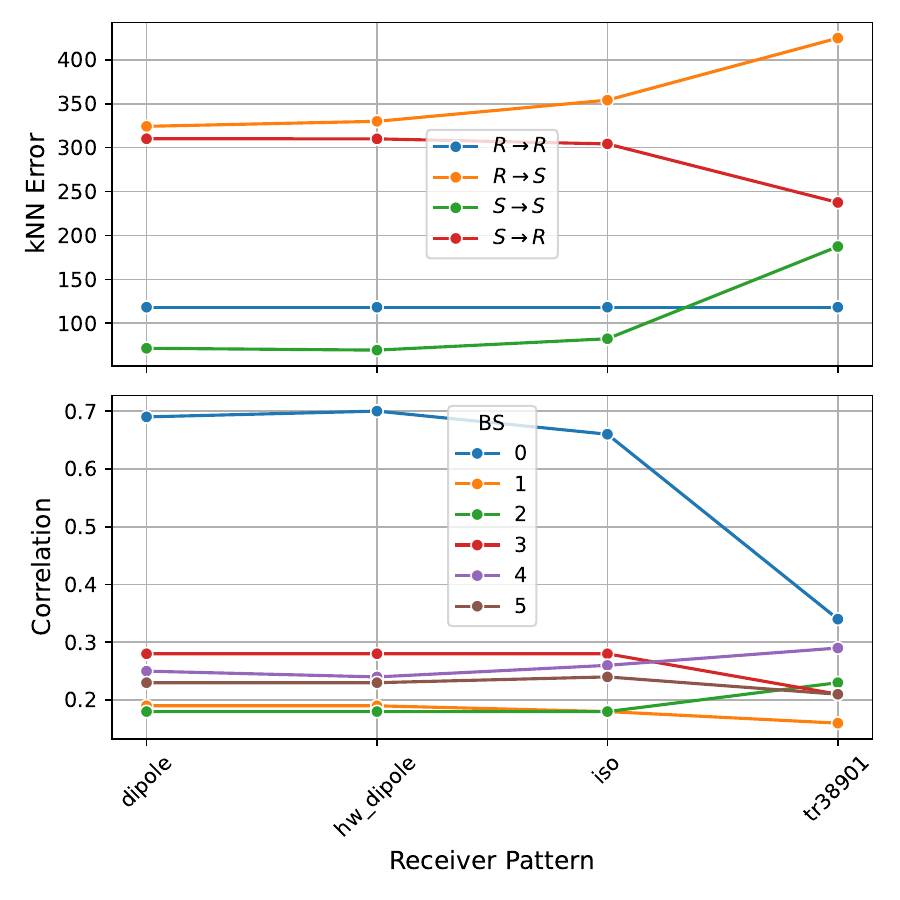}
    \caption{UE radiation pattern}
    \label{fig:rx-pattern}
  \end{subfigure}\hfill
  \vspace{1em} 

  \begin{subfigure}[b]{0.32\textwidth}
    \centering
    \includegraphics[width=\linewidth]{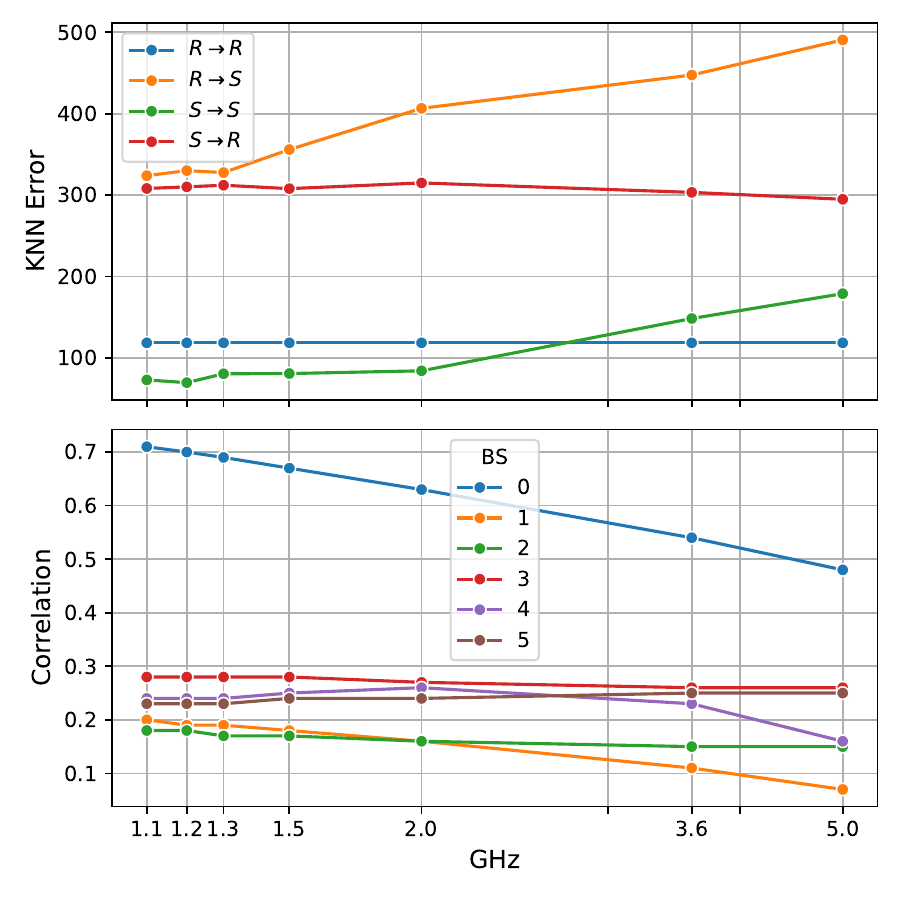}
    \caption{Frequency}
    \label{fig:frequency}
  \end{subfigure}\hfill
  \begin{subfigure}[b]{0.32\textwidth}
    \centering
    \includegraphics[width=\linewidth]{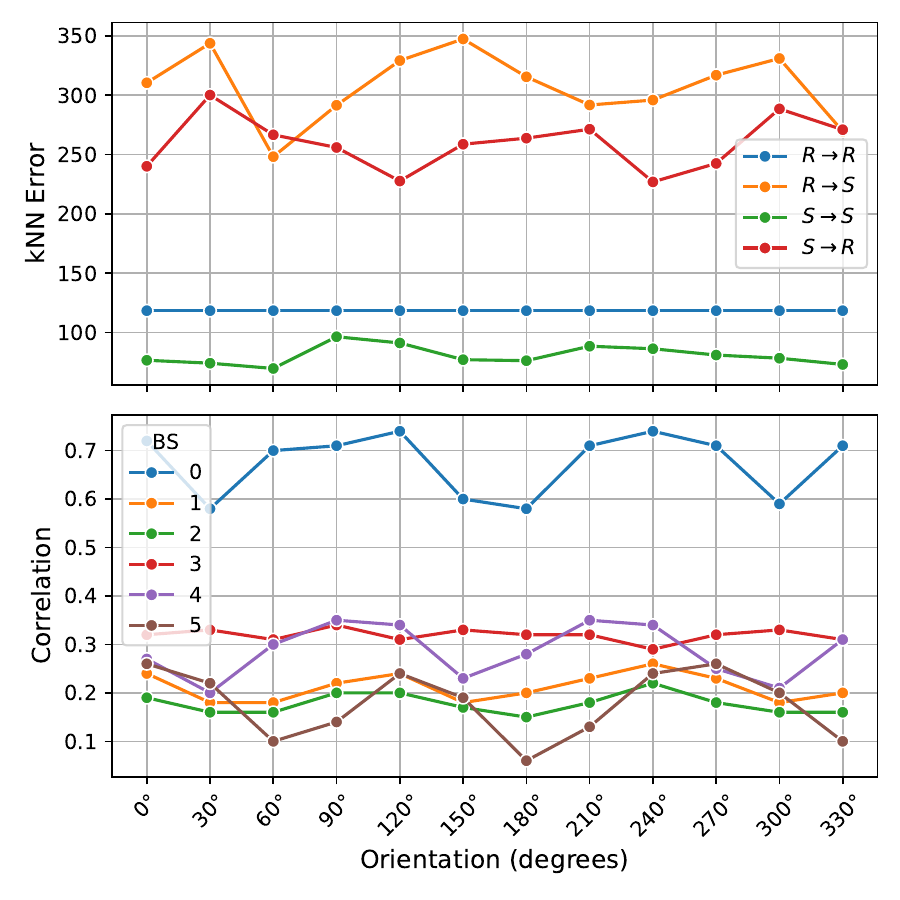}
    \caption{BS Orientation}
    \label{fig:orientation}
  \end{subfigure}\hfill
  \begin{subfigure}[b]{0.32\textwidth}
    \centering
    \includegraphics[width=\linewidth]{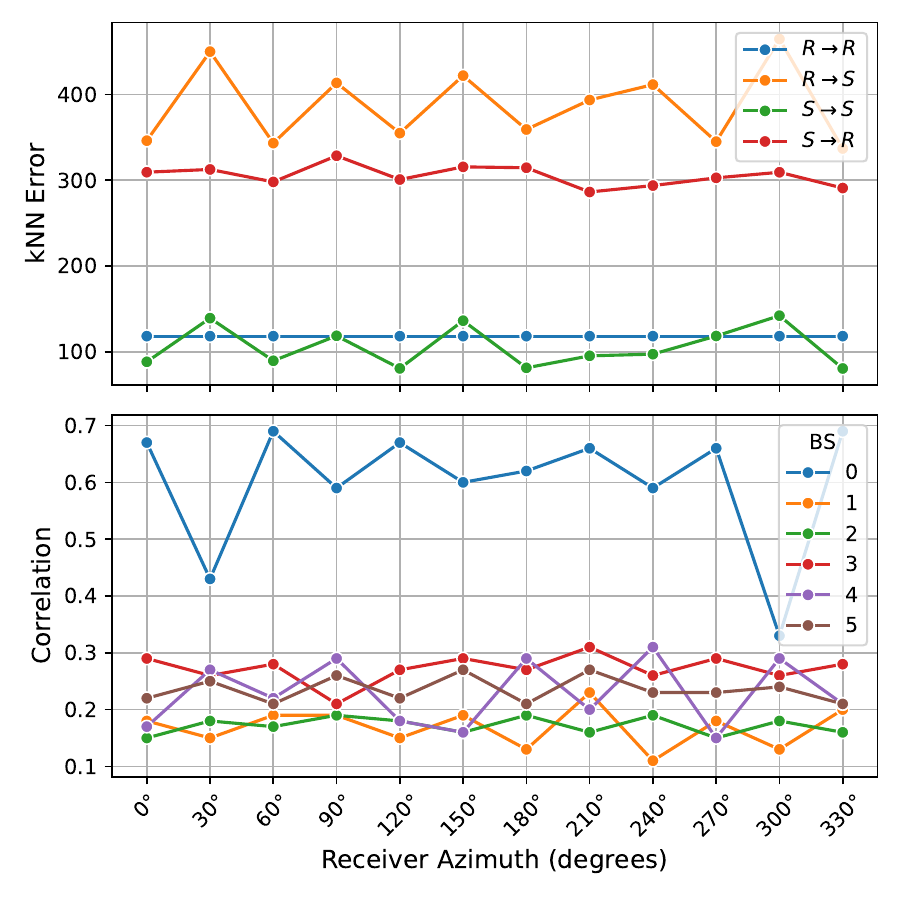}
    \caption{UE Orientation}
    \label{fig:rx-orientation}
  \end{subfigure}

  \caption{Sensitivity to various parameters of the simulated base stations and user equipments.}
  \label{fig:parameter-sweeps}
\end{figure*}

\begin{table*}
    \centering
        \caption{PathSolver parameters in Sionna v1.0.2, their default values, and the values we used in all experiments in this work.}

\begin{tabular}{llcc}
\toprule
Variable & Parameter name & Sionna default & Value in our experiments\\
\midrule
Maximum paths per source      & \texttt{max\_num\_paths\_per\_src} & $10^{6}$ & $10^{4}$\\
Samples per source            & \texttt{samples\_per\_src}         & $10^{6}$ & $10^{6}$\\
Maximum depth                 & \texttt{max\_depth}                & 3        & 3\\
Synthetic array               & \texttt{synthetic\_array}          & \texttt{True}  & \texttt{False}\\
Specular reflection           & \texttt{specular\_reflection}      & \texttt{True}  & \texttt{False}\\
Diffuse reflection            & \texttt{diffuse\_reflection}       & \texttt{False} & \texttt{True}\\
Refraction                    & \texttt{refraction}                & \texttt{True}  & \texttt{True}\\\midrule
Frequency & \texttt{frequency} & 3.6 GHz & 1.2 GHz \\
\bottomrule
\end{tabular}

    \label{tab:sionna-variables}
\end{table*}

\subsection{Sensitivity to the simulation parameters}

Sionna v1.0.2 offers a \texttt{PathSolver} class which performs the heavylifting of the computation. It has a certain set of simulation-wide parameters that control the quality of the simulation. Three of these parameters, maximum number of paths per source (\texttt{max\_num\_paths\_per\_src}), samples per source (\texttt{samples\_per\_src}), maximum depth (\texttt{max\_depth}), are numeric, and larger numbers correspond to higher accuracy. Four more settings: synthetic array, specular reflection, diffuse reflection and refraction are boolean.

For numeric variables we tried several values on the logarithmic scale ($10^3$, $10^4$, $10^5$, $10^6$) and small integers for the \texttt{max\_depth} parameter. For the maximum number of paths per source, the results with $10^4$ were different from $10^3$, while larger values gave identical results to $10^4$. On the other hand, larger numbers increased computation time, so we choose the smallest value that was consistent with the larger ones: $10^4$.

The results were more sensitive to \texttt{samples\_per\_src} variable. We kept the default value of $10^6$, as the lower values gave significantly different results at least for some base stations (Fig. \ref{fig:samples-per-src}). 

Surprisingly, the results were absolutely insensitive to the \texttt{max\_depth} parameter. We kept the default value $3$ for our experiments.

We tried both boolean values for all four configuration variables: synthetic array (\texttt{synthetic\_array}, it simplifies the calculation for antenna arrays), specular reflection (\texttt{specular\_reflection})
diffuse reflection (\texttt{diffuse\_reflection})
and refraction (\texttt{refraction}). The results were identical. Table \ref{tab:sionna-variables} shows solver configuration variables and the corresponding values. 

As seen in Fig. \ref{fig:samples-per-src}, the RSSI values are relatively well correlated with the values captured in real-world experiments only for the first base station. The other transmitters are correlated quite poorly. The rest of this paper is devoted to understanding whether certain simulation parameters are responsible for these poor results. 

\begin{figure*}
    \centering
    \includegraphics[width=0.7\linewidth]{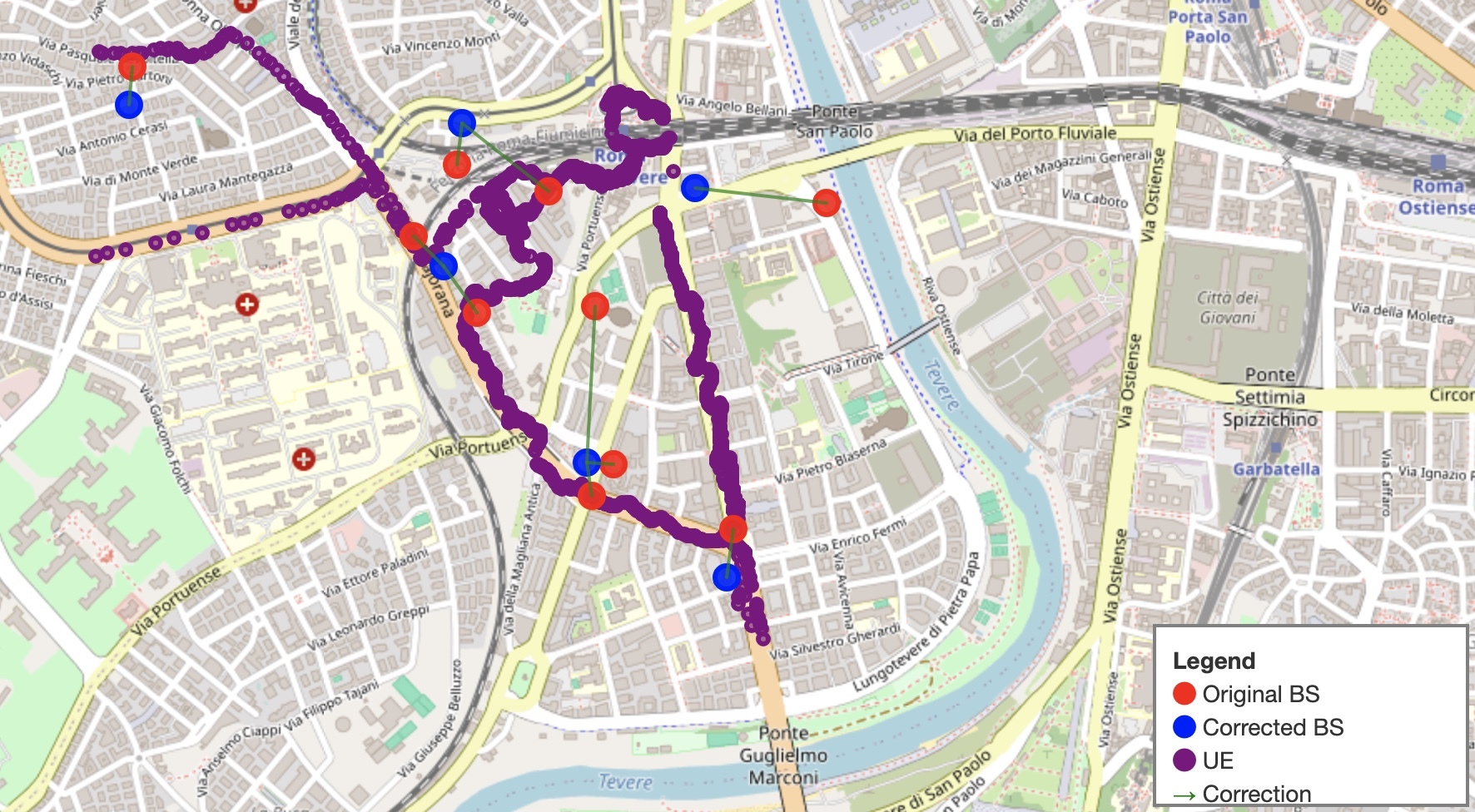}
    \caption{The dark purple and red points represent, respectively, user equipment and base station locations on the Rome map, as reported in \cite{RomeData}. The blue points indicate corrected locations. See more in Section \ref{sec:correcting-bs-locations}.}
    \label{fig:corrected-map}
\end{figure*}

\subsection{Correcting the BS locations} \label{sec:correcting-bs-locations}
Manual inspection highlighted a potentially critical issue. We could not visually find the base stations on high quality aerial imagery of Rome at the designated locations. We then used Google Earth imagery to manually look for base stations on top of the buildings near the reported locations and made corrections. Fig. \ref{fig:corrected-map} shows the original and corrected locations of base stations. 

Note that some of the corrections are quite far which seems suspicious. Also, note that in one case we could find only one correct location for three different BSs, and in another case one location is the closest one to two other BSs. To avoid confusion, we decided to remove these potentially ``duplicated'' transmitters from our data, and ended up with only $M=6$ base stations. All original UEs were kept. 

The change in the metrics was not significant, but nevertheless we decided to use this smaller and more accurate dataset for further experiments.

\subsection{Altitude of the base stations}
While most of the buildings were created to have 10m height, the optimal altitude of base stations was found to be non-trivial. We tried 11m, 12m, 15m, 20m, 35m, 40m and 55m altitudes for all transmitters, and noticed that the Spearman correlation is significantly better when the antennas are higher (Fig. \ref{fig:altitude}). It can also be seen that the error of kNN in the S $\rightarrow$ S scenario is higher than in the R $\rightarrow$ R scenario when the antenna is too low, which is another evidence that such simulations are unreliable. We used 40m altitude for all base stations for the rest of our experiments. Note that the most optimal configuration varies per base station. Fig. \ref{fig:altitude} shows BS \#2 performs best at 55m, while BS \#1 and \#4 are better at 40m altitude.

\begin{table*}[h!]
\centering
\caption{Spearman correlation between real and simulated RSSI values for each of the base stations in the initial and the final, optimized simulations. The last four columns show kNN localization errors in the four scenarios described in Section \ref{sec:eval}}
\label{tab:final-results}
\begin{tabular}{@{}lcccccccccc@{}}
\toprule
& BS \#0        & BS \#1        & BS \#2        & BS \#3        & BS \#4        & BS \#5        & R $\rightarrow$ R & R $\rightarrow$ S & S $\rightarrow$ S & S $\rightarrow$ R \\ \midrule
Spearman correlation @ initial configuration (11m alt.) & 0.54          & 0.11          & 0.15          & 0.26          & 0.23          & \textbf{0.25} & 118.32            & 447.49            & 148.21            & 303.26            \\
Spearman correlation @ optimized configuration          & \textbf{0.74} & \textbf{0.26} & \textbf{0.27} & \textbf{0.34} & \textbf{0.35} & \textbf{0.26} & 118.32            & 295.85            & 86.23             & 226.86            \\
Altitude in the optimized configuration                 & 40m           & 40m           & 55m           & 40m           & 40m           & 40m           &                   &                   &                   &                   \\
Azimuth in the optimized configuration                  & 120$^\circ$   & 240$^\circ$   & 0$^\circ$     & 90$^\circ$    & 210$^\circ$   & 0$^\circ$     &                   &                   &                   &                   \\ \bottomrule
\end{tabular}
\end{table*}

\subsection{The impact of the radio frequency} The dataset does not contain information on which 4G bands were used when collecting this data. We assume all measurements are performed using the same frequency. We performed a series of experiments and noticed that lower frequencies around 1GHz give the best correlations and the best kNN errors (Fig. \ref{fig:frequency}). Note that for higher frequencies than 2GHz, S $\rightarrow$ S scenario gives a larger error than the R $\rightarrow$ R scenario, which indicates that simulated RSSI values at higher frequencies are less suitable for fingerprinting.

We used $1.2$GHz as the main frequency for our subsequent experiments. 


\subsection{Radiation patterns}
Sionna supports four antenna patterns: isotropic (\texttt{iso}), two kinds of dipole (\texttt{dipole}, \texttt{hw\_dipole}) and a directional system defined in 3GPP TR 38.901 (\texttt{tr38901}). We tried experiments by setting all transmitters to have the same radiation pattern. 
While the last pattern is highly sensitive to antenna's orientation, it performed well with the best base station (\#3) under an arbitrarily chosen orientation (Fig. \ref{fig:pattern}). 
Also, RSSI fingerprints from Sionna act as pretty good features for kNN localization under tr38901. We chose it as the main pattern for the subsequent experiments.

We performed a similar experiment with receivers. Dipole-like patterns worked well both in terms of Spearman correlation with the real-world data on every base station, but also gave the lowest kNN error in the R $\rightarrow$ S scenario. We chose \texttt{hw\_dipole} for the subsequent experiments.

\subsection{Orientation of the devices}
Finally, we analyze the role of antenna orientations. We set tilt to -5 degrees and roll to 0 degrees to simplify the search space, and vary azimuth in the range of $[0, 330]$ with $30$ degree steps. 

As seen in Fig. \ref{fig:orientation}, each base station has one or more ``preferred'' orientations. BS \#0 performs best at 120 degree and 240 degree positions, while BS \#4 performs better at 90 and 210 degrees. Nevertheless we kept 0 transmitter azimuth for the subsequent experiments.

Similar to the base stations, we perform a small grid search for UE orientation as well. Fig. \ref{fig:rx-orientation} shows there is a significant variance, even if we enforce all receivers to have the same orientation. One could hope that these random orientations with respect to the antennas would cancel out the net effect, but even the kNN errors show significant variance depending on the receiver's orientation.

\subsection{Analysis}
Our extensive simulations highlighted a few patterns on the sensitivity of Sionna-based wireless simulations. Most of the scene-wide parameters of the solvers do not have a profound impact on the results. The highest sensitivity is observed from the radiation patterns and orientations of all devices. The latter is somewhat unfortunate, as it is impractical to control for the orientation of a mobile device used in an urban environment, and hence there is an inherent noise in the real data at least because of the random orientation.

The basic optimizations, mostly in BS orientation and radiation patterns, brought significant improvements in both correlations and kNN error rates. Table \ref{tab:final-results} shows the change from the initial configuration to the most optimized one we found. With the right configuration it was possible to improve upon the baseline by 50\% and even 2.3x in one case. Moreover, in the downstream application scenario it is possible to decrease the kNN localization error by roughly one-third by solely optimizing the simulation details. Still, after these optimizations, simulated data is not an immediate replacement for real-world data.


\subsection{Limitations of this work} 
A few aspects of the simulations are missing in this work. First, we used homogenous materials across the whole scene. The sensitivity of the RSSI signals with respect to the materials of the building is not explored yet. 
Another major gap between simulated and real environments is the granularity of the buildings and the environment in general. We were not able to find easy-to-use 3D models of buildings in Rome that are supported by Sionna's engine. We believe the complicated shapes of the buildings, exterior details, balconies, other architectural elements might have significant impact on radio signal propagation that is not captured with simplistic models we work with.

\section{Conclusions and future work} \label{sec:con}
In this work we performed roughly 400 simulations in a carefully selected region of Rome, for which there exists a publicly available dataset of real-world 4G/5G singal strengths. We showed that while simulated data is not well correlated with the real RSSI values, it is possible to significantly decrease the gap by carefully tuning certain aspects of the simulation, particularly the radiation patterns and orientations of the antennas. 

\paragraph{Recommendations for simulated data generation with Sionna}
Our experiments show that the solver's numerical and boolean parameters do not significantly affect the outcomes unless they are configured to use too little compute. Also, in addition to the locations of base stations, radiation patterns and orientations play a critical role in determining the signal strength. Somewhat surprisingly, higher frequencies tend to be less reliable for fingerprinting for localization purposes.

\paragraph{Future work} More detailed optimization of the simulation parameters might bring to more improvements in the correlation between simulated and real-world data points. Certain aspects of simulations that were left out of the scope of this work need more analysis (including the impact of the materials and more granular shapes of the buildings). Even if the simulation-to-real gap is not closed yet, the minimal costs of the simulations allow us to generate RSSI signal data for an extremely large number of points. Whether that can help in decreasing error rates in real-world localization problems is yet to be seen.

\section*{Acknowledgements}
This work was supported by funding under the bilateral agreement between CNR (Italy) and HESC MESCS RA (Armenia) as part of the DeepRF project for the 2025–2026 biennium, and by the HESC MESCS RA grant No. 22rl-052 (DISTAL).





\balance


\begin{thebibliography}{10}
\providecommand{\url}[1]{#1}
\csname url@samestyle\endcsname
\providecommand{\newblock}{\relax}
\providecommand{\bibinfo}[2]{#2}
\providecommand{\BIBentrySTDinterwordspacing}{\spaceskip=0pt\relax}
\providecommand{\BIBentryALTinterwordstretchfactor}{4}
\providecommand{\BIBentryALTinterwordspacing}{\spaceskip=\fontdimen2\font plus
\BIBentryALTinterwordstretchfactor\fontdimen3\font minus
  \fontdimen4\font\relax}
\providecommand{\BIBforeignlanguage}[2]{{%
\expandafter\ifx\csname l@#1\endcsname\relax
\typeout{** WARNING: IEEEtran.bst: No hyphenation pattern has been}%
\typeout{** loaded for the language `#1'. Using the pattern for}%
\typeout{** the default language instead.}%
\else
\language=\csname l@#1\endcsname
\fi
#2}}
\providecommand{\BIBdecl}{\relax}
\BIBdecl

\bibitem{9714331}
H.~A.~H. Alobaidy, M.~Jit~Singh, M.~Behjati, R.~Nordin, and N.~F. Abdullah,
  ``Wireless transmissions, propagation and channel modelling for iot
  technologies: Applications and challenges,'' \emph{IEEE Access}, vol.~10, pp.
  24\,095--24\,131, 2022.

\bibitem{9682024}
A.~Elmaghbub and B.~Hamdaoui, ``Comprehensive rf dataset collection and
  release: A deep learning-based device fingerprinting use case,'' in
  \emph{2021 IEEE Globecom Workshops (GC Wkshps)}, 2021, pp. 1--7.

\bibitem{Yun2024}
\BIBentryALTinterwordspacing
Z.~Yun and M.~F. Iskander, \emph{Radio Propagation Modeling and Simulation
  Using Ray Tracing}.\hskip 1em plus 0.5em minus 0.4em\relax Cham: Springer
  International Publishing, 2024, pp. 251--279. [Online]. Available:
  \url{https://doi.org/10.1007/978-3-031-39824-7\_10}
\BIBentrySTDinterwordspacing

\bibitem{KHACHATRIAN2025103696}
\BIBentryALTinterwordspacing
H.~Khachatrian, R.~Mkrtchyan, and T.~P. Raptis, ``Deep learning with synthetic
  data for wireless nlos positioning with a single base station,'' \emph{Ad Hoc
  Networks}, vol. 167, p. 103696, 2025. [Online]. Available:
  \url{https://www.sciencedirect.com/science/article/pii/S157087052400307X}
\BIBentrySTDinterwordspacing

\bibitem{10592367}
------, ``Outdoor environment reconstruction with deep learning on radio
  propagation paths,'' in \emph{2024 International Wireless Communications and
  Mobile Computing (IWCMC)}, 2024, pp. 1498--1503.

\bibitem{sionna-rt}
J.~Hoydis, F.~A. Aoudia, S.~Cammerer, M.~Nimier-David, N.~Binder, G.~Marcus,
  and A.~Keller, ``Sionna rt: Differentiable ray tracing for radio propagation
  modeling,'' in \emph{2023 IEEE Globecom Workshops (GC Wkshps)}, 2023, pp.
  317--321.

\bibitem{RomeData}
K.~Kousias, M.~Rajiullah, G.~Caso, U.~Ali, O.~Alay, A.~Brunstrom, L.~De~Nardis,
  M.~Neri, and M.-G. Di~Benedetto, ``A large-scale dataset of 4g, nb-iot, and
  5g non-standalone network measurements,'' \emph{IEEE Communications
  Magazine}, vol.~62, no.~5, pp. 44--49, 2024.

\bibitem{10827549}
Y.~J. Noh and K.~W. Choi, ``High-precision digital twin platform based on ray
  tracing simulation,'' in \emph{2024 15th International Conference on
  Information and Communication Technology Convergence (ICTC)}, 2024, pp.
  1464--1465.

\bibitem{Schott2023EUCAP}
A.~Schott, A.~Ichkov, P.~M{\"a}h{\"o}nen, and L.~Simi{\'c}, ``Measurement
  validation of ray-tracing propagation modeling for mm-wave networking
  studies: How detailed is detailed enough?'' in \emph{Proc. 17th European
  Conf. on Antennas and Propagation (EuCAP)}, 2023, pp. 1--5.

\bibitem{Di2024EUCAP}
J.~Di, Z.~Yuan, Y.~Lyu, F.~Zhang, and W.~Fan, ``Validation of ray-tracing
  simulated channels for massive {MIMO} systems at millimeter-wave bands,'' in
  \emph{Proc. 18th European Conf. on Antennas and Propagation (EuCAP)}, 2024,
  pp. 1--5.

\bibitem{Aksoy2024THzRT}
E.~Aksoy, A.~Schultze, A.~Fazli, L.~Raschkowski, L.~Azpilicueta,
  M.~Celaya-Echarri, M.~Navarro-C{\'i}a, and S.~Stanczak, ``Analysis of
  interaction mechanisms and intercomparison of ray‑tracing tools for
  optimizing thz simulations,'' in \emph{Proc.\ 2024 IEEE Global Communications
  Conference (GLOBECOM)}, 2024, pp. 1--6.

\bibitem{10666097}
G.~Xia, C.~Zhou, F.~Zhang, Z.~Cui, C.~Liu, H.~Ji, X.~Zhang, Z.~Zhao, and
  Y.~Xiao, ``Path loss prediction in urban environments with sionna-rt based on
  accurate propagation scene models at 2.8 ghz,'' \emph{IEEE Transactions on
  Antennas and Propagation}, vol.~72, no.~10, pp. 7986--7997, 2024.

\bibitem{sionna102}
F.~A. Aoudia, J.~Hoydis, M.~Nimier-David, S.~Cammerer, and A.~Keller, ``Sionna
  rt: Technical report,'' \emph{arXiv preprint arXiv:2504.21719}, 2025.

\end{thebibliography}
\end{document}